\documentclass[amsmath,amssymb, aps, prx, twocolumn,superscriptaddress,nofootinbib]{revtex4-2}

\usepackage{float}
\usepackage{graphicx}% Include figure files
\usepackage{dcolumn}% Align table columns on decimal point
\usepackage{bm}% bold math
\usepackage{hyperref}% add hypertext capabilities
\hypersetup{
    colorlinks=true,
    linkcolor=blue,
    filecolor=magenta,      
    urlcolor=cyan,
}
\usepackage{color}
\usepackage{fancyhdr}
\usepackage[normalem]{ulem}
\usepackage{physics}
\usepackage[dvipsnames]{xcolor}
\usepackage{gensymb}

\newcommand{\resp}[1]{\textcolor{black}{ #1 }}

\def\PA{School of Physics and Astronomy, University of Birmingham, Edgbaston, Birmingham, B15 2TT, United Kingdom}
\def\ENS{Physics Department, Ecole Normale Supérieure, PSL University, 24 rue Lhomond, 75005 Paris, France}

\begin{document}
%\preprint{APS/123-QED}
\title{Controlling the interactions in a cold atom quantum impurity system}

\author{Thomas Hewitt}
\affiliation{\PA}
\author{Tom Bertheas}
\affiliation{\PA}
\affiliation{\ENS}
\author{Manan Jain}
\affiliation{\PA}
\author{Yusuke Nishida}
\affiliation{Department of Physics, Tokyo Institute of Technology,
Ookayama, Meguro, Tokyo 152-8551, Japan}
\author{Giovanni Barontini}
\email{g.barontini@bham.ac.uk}
\affiliation{\PA}

\date{\today}

\begin{abstract}
We implement an experimental architecture in which a single atom of K is trapped in an optical tweezer, and is immersed in a bath of Rb atoms at ultralow temperatures. In this regime, the motion of the single trapped atom is confined to the lowest quantum vibrational levels. This realizes an elementary and fully controllable quantum impurity system. For the trapping of the K atom, we use a species-selective dipole potential, that allows us to independently manipulate the quantum impurity and the bath. We concentrate on the characterization and control of the interactions between the two subsystems. To this end, we perform Feshbach spectroscopy, detecting several inter-dimensional confinement-induced Feshbach resonances for the KRb interspecies scattering length, that parametrizes the strength of the interactions. We compare our data to a theory for inter-dimensional scattering, finding good agreement. Notably, we also detect a series of p-wave resonances stemming from the underlying free-space s-wave interactions. We further determine how the resonances behave as the temperature of the bath and the dimensionality of the interactions change. Additionally, we are able to screen the quantum impurity from the bath by finely tuning the wavelength of the light that produces the optical tweezer, providing us with a new effective tool to control and minimize the interactions. Our results open a range of new possibilities in quantum simulations of quantum impurity models, quantum information, and quantum thermodynamics, where the interactions between a quantized system and the bath is a powerful yet largely underutilized resource.

\end{abstract}

\maketitle

\section{Introduction}

A quantum impurity system is composed of a subsystem with a few degrees of freedom, the \emph{impurity}, interacting with a larger subsystem with potentially infinite degrees of freedom, the \emph{bath}. This class of systems is represented by the three-term Hamiltonian
\begin{equation} \label{eq:1}
H=H_I+H_B+H_{IB},
\end{equation}
where $H_I$ and $H_B$ are the impurity and bath Hamiltonians respectively, and $H_{IB}$ describes the coupling between the impurity and the bath. In general, $[H_I,H_{IB}]\neq0$. Several fundamental quantum mechanical models are based on this apparently simple construction, the most celebrated being the Anderson impurity model \cite{anderson1961localized,evers2008anderson}, the Kondo model \cite{kondo1964resistance,wilson1975renormalization} and the spin-boson model \cite{leggett1987dynamics}, whose impact and ramifications range from condensed matter physics to quantum information physics, the physics of open quantum systems, down to the very foundations of quantum mechanics. In the context of quantum technology, quantum dots are a prominent example of quantum impurity systems \cite{hanson2007spins,reimann2002electronic}, while cold atoms have been instrumental in implementing quantum simulators for quantum impurity models and their extensions \cite{bauer2013realizing,billy2008direct,roati2008anderson,schmidt2018quantum,schirotzek2009observation,zipkes2010trapped,fukuhara2013quantum}. 

{In quantum information science, significant efforts are taken to shield a quantum system, for example a qubit, from the environment. This helps reducing unwanted effects such as decoherence and dephasing. If properly controlled and tamed however, the interaction of a quantum system with the environment, embodied by $H_{IB}$, can be a powerful resource. For example, controlling the interaction of qubits with the environment could lead to the realization of environmental dark states \cite{verstraete2009quantum,muller2012engineered} and environmental resilient quantum information protocols \cite{goold2016role,alicki2004thermodynamics,loss1998quantum}. The control of the interaction of a quantum system and a bath is a central ingredient in quantum thermodynamics \cite{doi:10.1080/00107514.2016.1201896} and for the realization of quantum engines, see e.g. \cite{quan2007quantum, barontini2019ultra,rossnagel2016single,bouton2021quantum}. In these systems, quantum information is produced by the controlled exchange of heat with the environment. Coherent quantum engines can generate substantially more power than stochastic ones \cite{uzdin2015equivalence}. Employing non-thermal or non-classical baths can lead to more efficient and more powerful engines \cite{19}. Advanced transformations like the shortcut to adiabaticity can be used to improve the performances of quantum engines \cite{campo2014more}  and to generate engines assisted by a Maxwell’s demon \cite{quan2007quantum}. The bath itself could in turn be used to mediate the interaction between distant quantum systems or entangle them \cite{PhysRevLett.89.277901}. Emerging questions concerning the management of energy in the quantum regime must also be addressed understanding how a quantum system interacts with the environment \cite{vinjanampathy2016quantum,auffeves2022quantum}.}

Cold atom technology provides us with the largest set of tools to engineer and control the three terms in Eq.\ (\ref{eq:1}). In particular, optical tweezers and optical lattices allow us to confine the motion of ultracold atoms to a few degrees of freedom \cite{kaufman2012cooling,thompson2013coherence,RevModPhys.78.179,RevModPhys.80.885}, mesoscopic traps can be used to control the density and temperature of the atoms in an ultracold bath, and Feshbach resonances enable the control of the interactions \cite{chin2010feshbach,kohler2006production}. Here, we combine all these elements and realize a cold atom quantum impurity architecture by trapping a single K atom in an optical tweezer, and immersing it in a bath of Rb atoms at ultracold temperatures. To obtain the selective trapping of the K atom, we utilize a species-selective tweezer (SST), whose wavelength is tuned at the so-called tune-out wavelength for Rb, so that it is almost transparent for this atomic species \cite{catani2009entropy,sheng2022defect}. 

The main focus of this work is on the control of the coupling term $H_{IB}$ in Eq.\ (\ref{eq:1}). In our system, the coupling stems from ultracold collisions between the single atom impurity and the atoms in the bath. We show that, due to the mixed dimensionality of the system, the effective scattering length that parametrizes such interactions exhibits series of higher partial wave resonances arising from the underlying s-wave collisions \cite{massignan2006,nishida2010confinement}. By changing the external magnetic field, we detect several resonances in the 1D-3D regime. We compare our results with a theory for inter-dimensional scattering, finding very good agreement. This theory allows us to determine the inter-species scattering length as a function of the external magnetic field, providing us with the control of the interactions between the single atom quantum impurity and the bath, including the switching between resonant s-wave and p-wave collisions. By decreasing the temperature of the bath, we bring the impurity close to the motional ground state, further changing the dimensionality of the collisions to 0D-3D. By using a simplified description that accounts for the intrinsic anisotropy of the confinement produced by the tweezer, we show that new series of resonances appear that substantially change the scattering scenario.  

To achieve a higher degree of control, we exploit the wavelength of the SST to screen the single atom from the bath. In fact, we show that by red detuning the SST from the tune-out wavelength, the potential exerted on the Rb atoms becomes increasingly repulsive, preventing the Rb atoms from occupying the volume of the SST, and reducing the scattering rate up to a point where $H_{IB}\simeq0$. We further show that, in our system, $H_{IB}\neq0$ implies that $H$ is actually non-hermitian, because the atom-bath interactions lead to losses in both subsystems due to three-body collisions. We characterize the interplay between hermitian and non-hermitian dynamics, that respectively lead to thermalization and losses, and find that there are optimal regimes for thermalization.    

Our results, together with the established techniques to control $H_I$ and $H_B$, provide new capabilities in quantum simulations of quantum impurity models {and in quantum information science. In particular, our work delivers a crucial milestone in quantum thermodynamics towards the realization of single atom engines in the quantum regime \cite{quan2007quantum, barontini2019ultra}}. 

\section{Realization of a K quantum impurity in a Rb thermal bath}

\begin{figure}
 \centering
 \includegraphics[width=0.49\textwidth]{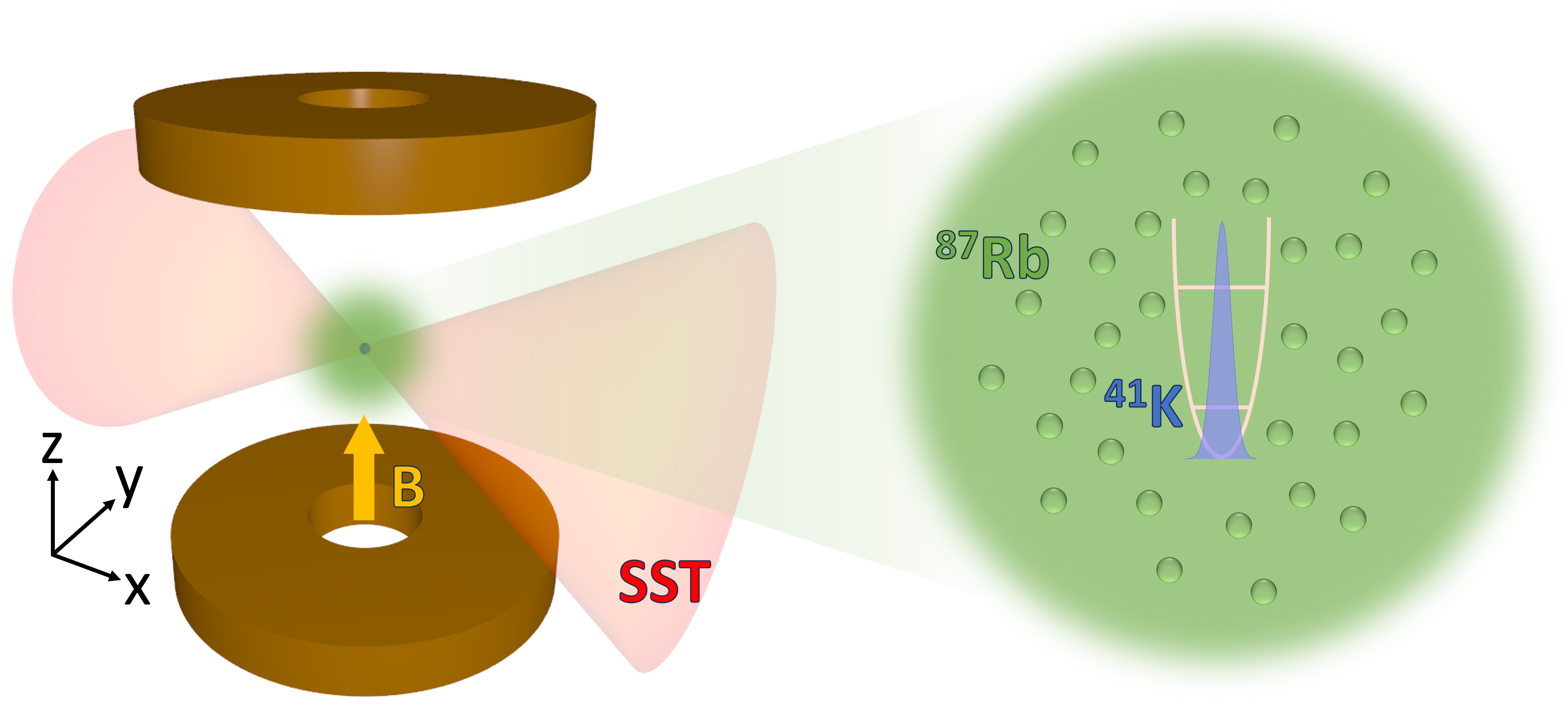}
 \caption{Representation of our experimental architecture (not to scale): a single $^{41}$K atom is trapped in a species-selective optical tweezer (SST) and immersed in a cloud of ultracold $^{87}$Rb atoms trapped in an optical dipole trap (not shown). Because of the tight confinement, in the ultracold regime the motion of the K atom is limited to a few quantized modes. A uniform magnetic field is applied along the vertical axis to control the value of the interspecies scattering length. By adjusting the wavelength of the SST around $\lambda_0$, we can make the tweezer attractive, transparent, or repulsive for the Rb atoms, allowing us to locally control the Rb density in the vicinity of the K atom.}
 \label{Fig1}
 \end{figure}

Our experimental architecture is sketched in Fig.\ \ref{Fig1}: a single atom of $^{41}$K is trapped in a SST and is immersed in a cloud of $^{87}$Rb at ultracold temperatures. The wavelength of the laser that produces the SST is tuned close to the tune-out wavelength for Rb \cite{catani2009entropy,schmidt2016precision}, $\lambda_0\simeq790$ nm. An external magnetic field is used to control the interspecies interaction strength. This configuration realizes a pristine implementation of a quantum impurity system, because at ultralow temperatures, the motion of the tightly trapped K atom is confined to a few modes of a three-dimensional quantum harmonic oscillator.   

Our experimental sequence starts by loading Rb atoms from a dark magneto-optical trap (MOT) in a crossed dipole trap \cite{munoz2020dissipative}. The atoms are initially pumped in the $|F=1\rangle$ ground state manifold. We then perform evaporative cooling keeping the magnetic quadrupole field on, so that at the end of the evaporation we obtain a polarized sample in the $|F=1,m_F=1\rangle$ state. A typical evaporation ramp lasts 4.25 s, and we obtain pure Bose-Einstein condensates of 6$\times$10$^{4}$ atoms. Unless otherwise stated, for this work, we stop the evaporation after 2.25~s obtaining thermal gases of 10$^{5} $ atoms at $\simeq1$ $\mu$K, with trapping frequencies $\omega^{(Rb)}=2\pi\times$ (85, 255, 265) Hz.  At the end of the ramp, we switch on a K MOT that surrounds the Rb sample for 200 ms. At the same time we switch on the SST. The latter is realized with an apoplanar microscope objective that produces a beam with $\simeq2.1$ $\mu$m waist at a distance of 25.5 mm. The SST light is linearly polarized along the vertical direction. The SST is initially placed 40 $\mu$m above the Rb cloud to avoid losses due to KRb scattering during the loading of the K atoms. The trapping potential has a typical depth of 1.2 mK. After the MOT, we compress the K cloud with a dark MOT, allowing us to load $\simeq$ 4 atoms of K in the SST with temperatures of 30 $\mu$K. We then switch off the quadrupole field and we ramp on a uniform magnetic field of 10 G along the vertical direction. Once the field is set, we optically pump the K atoms in the $|F=1,m_F=1\rangle$ state with two laser beams tuned on the D1 line. The beams include a $\pi$ polarized beam resonant with the $|2S_{1/2}, F=2 \rangle \rightarrow |2P_{1/2}, F=2 \rangle$ transition and a $\sigma^{+}$ polarized beam resonant with the $|2S_{1/2}, F=1 \rangle \rightarrow |2P_{1/2}, F=1 \rangle$ transition.  After this stage, both species are therefore in the target $|F=1,m_F=1\rangle$ ground state. Finally, we ramp the magnetic field to the target value in 50 ms, lowering at the same time the power of the SST. This allows us to keep only the coldest atoms in the SST and to reduce the off-resonant light scattering that could cause losses for both species. Typically, we end up with 1.5 atoms of K at $\simeq$ 20 $\mu$K, with trapping frequencies $\omega^{(K)}=2\pi\times(1.4,22,22)$ kHz.   
\par
To immerse the K atoms in the Rb bath we actuate a mechanical nanpositioner stage that brings the SST to the centre of the Rb cloud in 50 ms. After a variable interaction time, the Rb atoms are released, and we perform time of flight absorption imaging. The K atoms are instead kept in the SST and their number is inferred through fluorescence imaging, typically needing 30 repetitions. In order to avoid undesired light shifts caused by the SST, we chop the fluorescence and tweezer light at a frequency of 1.02 MHz and a phase shift of 90$\degree$ \cite{hutzler2017eliminating}. To measure the temperature of the K atoms, we perform release-recapture thermometry using Bayesian estimation, requiring 180 experimental runs \cite{glatthard2022optimal}.

The tune-out wavelength $\lambda_0$ for $^{87}$Rb is approximately 790 nm \cite{schmidt2016precision} in case of linearly polarized light. In our case however, the strong focussing of the tweezer beam could in principle generate longitudinal field components at the position where the K atoms are trapped. This would result in a shift of the tune-out wavelength. Therefore, to find where the SST is transparent for the Rb atoms, we measure the survival probability of the K atoms as a function of the SST wavelength. Indeed, across $\lambda_0$ the potential for Rb goes quite steeply from attractive to repulsive, respectively increasing or decreasing the local Rb density. By scanning the SST wavelength we observe that the survival probability of K atoms at 10 G after 50 ms goes from 0 to 1 in the [790-790.1] nm range. This sets our range of operations as it allows us to fully open or close the SST volume to the Rb atoms. We will show in Sec. IV how to use this feature to control the way the quantum impurity interacts with the bath. We calculate that at 0.1 nm detuning from $\lambda_0$ the density of Rb in the SST volume is vanishing, therefore we infer that for our system $\lambda_0$ is not significantly shifted from the expected value of $\simeq790$ nm \cite{SuppMat}. 

\begin{figure}
 \centering
 \includegraphics[width=0.49\textwidth]{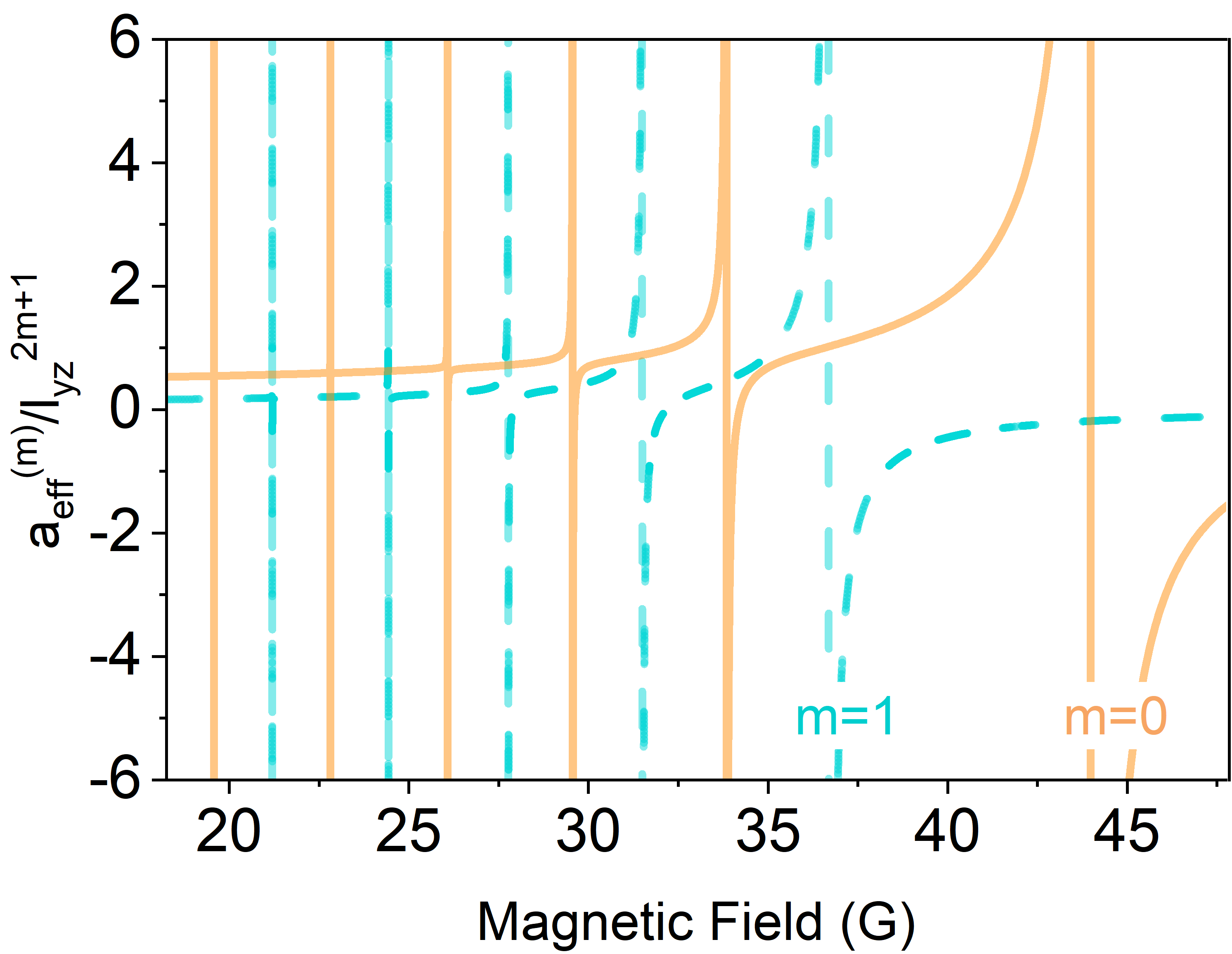}
 \caption{{Theoretical predictions for the effective scattering `length' in units of $l_{yz}^{2m+1}$ as a function of the external magnetic field for a 1D-3D system with our experimental parameters. The solid orange line corresponds to the s-wave scattering length ($m=0$), the dashed cyan line to the p-wave scattering volume ($m=1$).}}
 \label{Fig1th}
 \end{figure}

 \begin{figure*}
 \centering
 \includegraphics[width=0.99\textwidth]{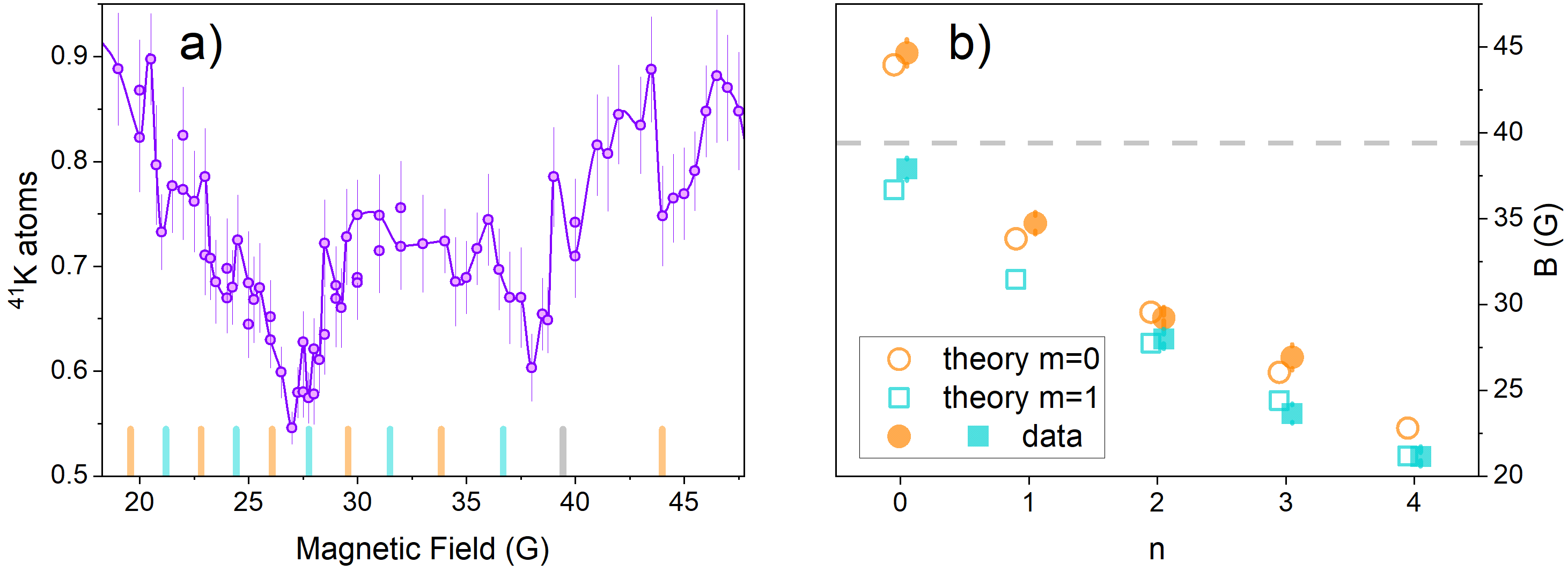}
 \caption{{a) Feshbach spectroscopy for our quantum impurity system in the 1D-3D regime (1 $\mu$K): number of remaining K atoms as a function of the external magnetic field after 50 ms of interaction. For each value of the magnetic field, we average over 30 experimental repetitions. The errorbars are the standard deviation of the mean. The solid purple line is a spline curve to guide the eye. The vertical grey line is the position of the free-space resonance according to \cite{simoni2008near}. The vertical orange (cyan) lines are the positions of the $m=0$ ($m=1$) resonances predicted by the theory. b) The open symbols are the values of the magnetic fields for the $m=0$ (circles) and $m=1$ (squares) inter-dimensional Feshbach resonances predicted by the theory, as a function of the order integer $n$ in Eq. (\ref{eqbinding}). The filled symbols are the magnetic field values of the centres of the loss features in a) obtained with a Gaussian fit. Data and theory are offset along the horizontal axis to enhance readability.  We append the widths of the resonances obtained with the fit as `errorbars'. We assign each experimental resonance to the closest theoretical one. The dashed line is the position of the 3D resonance.}}
 \label{Fig2}
 \end{figure*}

 \begin{figure}
 \centering
 \includegraphics[width=0.49\textwidth]{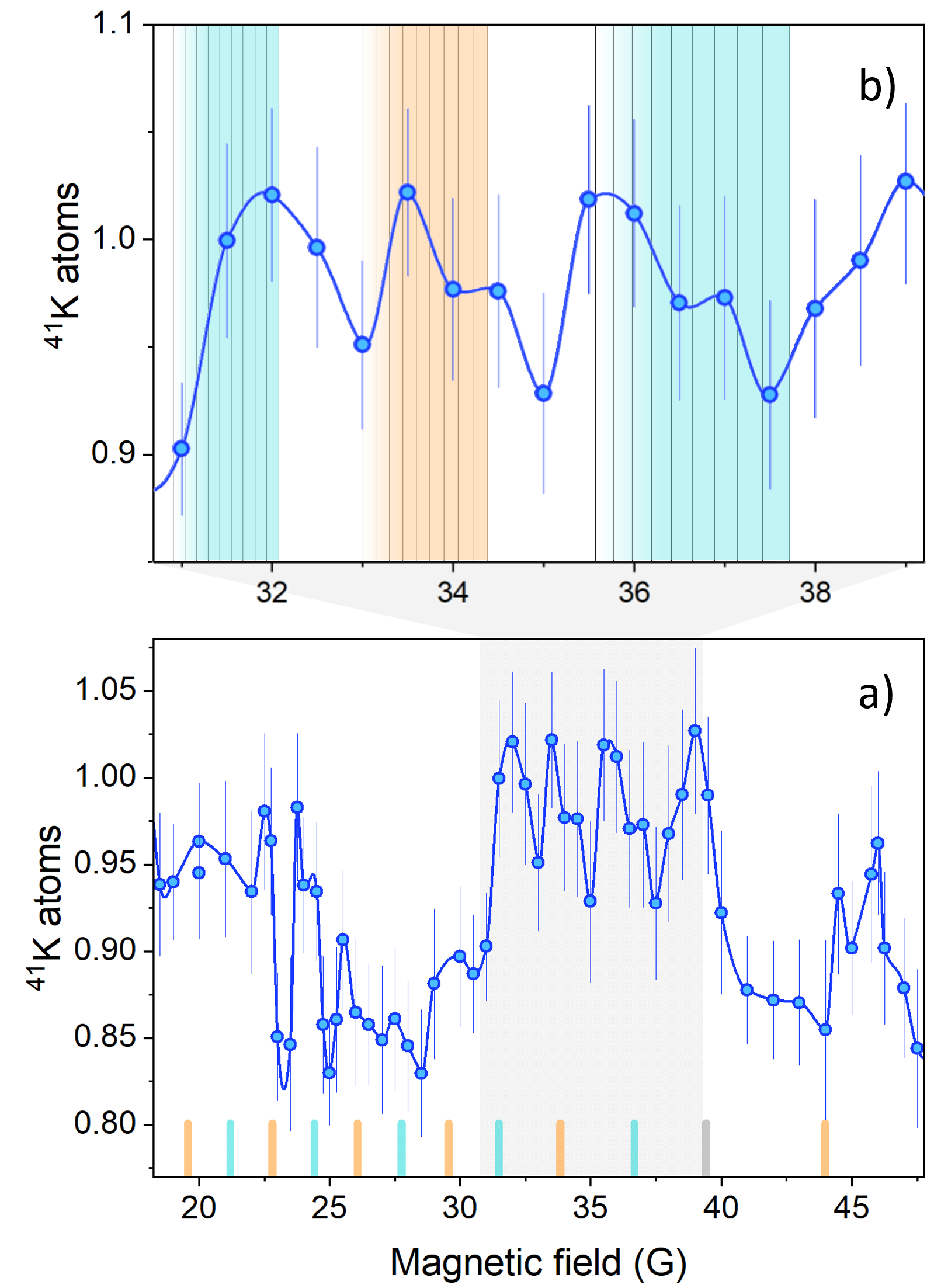}
 \caption{{a) Feshbach spectroscopy for our quantum impurity system in the 0D-3D regime  (100 nK): number of remaining K atoms as a function of the external magnetic field after 150 ms of interaction. For each data point we average over 30 experimental repetitions. The errorbars are the standard deviation of the mean. The solid blue line is a spline curve to guide the eye. The vertical grey line is the position of the free-space resonance according to \cite{simoni2008near}. The vertical orange (cyan) lines are the positions of the $m=0$ ($m=1$) resonances predicted by the theory. b) Zoom in the shaded grey area of a). The vertical black lines are the positions of the resonances according to Eq.\ (\ref{eqbinding2}) up to the tenth order. The shading indicates that the net effect is broadened resonances that are skewed towards the left, as observed in the experiment.}}
 \label{Fig22}
 \end{figure}

 \begin{figure*}
 \centering
 \includegraphics[width=0.99\textwidth]{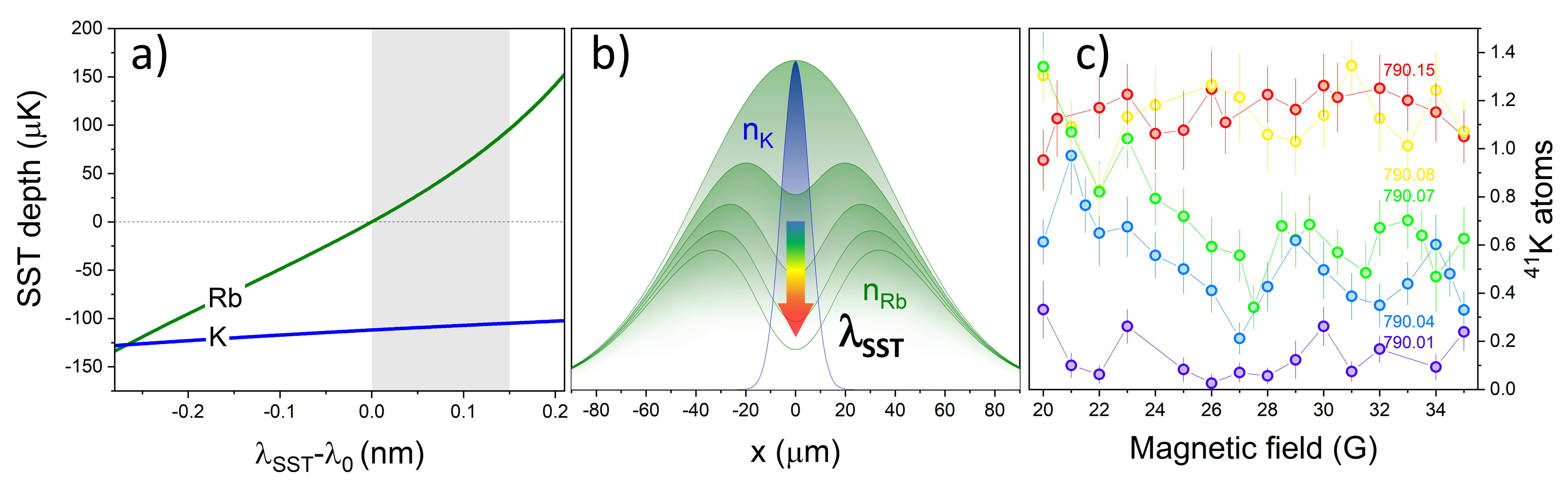}
 \caption{Controlling the interaction with the SST wavelength. a) For Rb, the SST potential is attractive (repulsive) at wavelengths below (above) \resp{$\lambda_0=$ 790 nm}. For K, the SST potential is always attractive and not significantly modified around $\lambda_0$. The grey shaded area indicates the wavelengths utilized in this work. b) By tuning the wavelength of the SST across $\lambda_0$, we can control the density of Rb in the tweezer volume. {The plotted density distributions are evaluated for different values of $\lambda_{SST}$ \cite{SuppMat}.} c) Number of K atoms remaining in the SST after 50 ms as a function of the magnetic field, for different values of $\lambda_{SST}$. The temperature of the Rb sample is 1 $\mu$K. Each data point is the average over 30 experimental repetitions, and the errorbars are the standard deviation of the mean. 
 }
 \label{Fig5}
 \end{figure*}

\section{Controlling the interactions with the external magnetic field}

In general, the $^{41}$K-$^{87}$Rb mixture is particularly appealing to control the interspecies interactions, because it features two fairly broad s-wave Feshbach resonances and two zero-crossings of the s-wave scattering length below 100 G \cite{thalhammer2008double}. However, if one of the two species is tightly trapped, the scenario becomes substantially richer. In particular, when $k_BT\leq\hbar\omega_{i}^{(K)}$, with $i=x,y,z$, the motion of the K atom in the $i$ direction(s) is `frozen', effectively confining the motion to the remaining dimension(s). Considering $\omega_{y}^{(K)}=\omega_{z}^{(K)}=:\omega_{yz}^{(K)}$ in our tweezer, by lowering the temperature it is possible to progressively reduce the dimensionality of the K atom motion from 3D to 1D to 0D. The motion of the Rb atoms in the bath is instead well described by the thermal motion of particles in a three dimensional harmonic potential with frequencies $\omega^{(Rb)}\ll\omega^{(K)}$. This means that the interaction of the K quantum impurity with the Rb bath effectively happens in mixed dimensions, substantially modifying the behaviour of the interspecies scattering length from the 3D-3D regime \cite{massignan2006,nishida2008,lamporesi2010scattering,nishida2010confinement}. In particular, it is expected that a single s-wave Feshbach resonance generates series of Feshbach resonances in s- and higher partial waves. \resp{This is because of the tight trapping of the K atoms that in turns
implies tight confinement of the Feshbach KRb molecule,
as well as coupling between the center-of-mass and relative motions.} In the 1D-3D regime with $\omega_{x}^{(K)}$ and $\omega^{(Rb)}$ neglected, new resonances are approximately expected to appear when the energy of the confined KRb molecule coincides with the scattering threshold at
\begin{equation}
   \left(1+|m|+2n\right)\sqrt{\frac{m_K}{m_K+m_{Rb}}}\hbar\omega^{(K)}_{yz}-|E_b| = \hbar\omega^{(K)}_{yz}
   \label{eqbinding}
\end{equation}
with $|m|$ the magnetic quantum number, $n\geq0$ an integer, $m_{K(Rb)}$ the K(Rb) mass, and $|E_b|$ the KRb binding energy in free space \cite{nishida2010confinement}. The 3D-3D resonance is instead expected to disappear. Therefore, for each $n$ there are series of resonances in higher partial waves, labelled with $m$. Note that, as evident in Eq.\  (\ref{eqbinding}), the positions of s-wave ($m=0$) and p-wave ($m=1$) resonances essentially overlap with those of d-wave ($m=2$) and f-wave ($m=3$) resonances respectively. Because higher order resonances are expected to be weaker, we will consider only $m=0$ and $m=1$ resonances. While confinement-induced resonances have been observed in \cite{haller2010confinement,lamporesi2010scattering,capecchi2022observation}, so far there hasn't been an unambiguous observation of higher partial wave resonances induced by mixed dimensional scattering. The ability to switch between resonant s- and p-wave interactions in a quantum impurity system is of great interest because it could be used to realize the SU(3) orbital Kondo effect, exploiting the threefold orbital degeneracy of p-wave resonances in the 0D-3D regime \cite{PhysRevLett.111.135301, PhysRevA.93.011606}. 

To accurately determine the effective scattering length $a_{eff}^{(m)}$ as a function of the external magnetic field and the resonance positions in our system, we utilize the theory developed in \cite{nishida2010confinement} with our experimental parameters. The regularized wave function of two particles at the same position projected to the $m$th partial wave $\chi_m(\rho)$ obeys the following integral equation derived from the Schr\"odinger equation:
\begin{align}
\frac1{\tilde a(\hat E_c)}\chi_m(\rho)
&= \frac{\rho^me^{-\rho^2/2l_{yz}^2}}{(2m+1)!\,\pi\,l_{yz}^2} \notag\\
&\quad + \frac{2\pi}{\mu_{KRb}}\int\!d^{2\!}\bm\rho'\,\mathcal G_m(\rho;\rho')\chi_m(\rho').
\end{align}
Here $l_{yz}=\sqrt{\hbar/m_K\omega^{(K)}_{yz}}$ is the transverse harmonic oscillator length, $\mu_{KRb}=m_{K}m_{Rb}/(m_K+m_{Rb})$ is the reduced mass, $\mathcal G_m(\rho;\rho')$ is the regular part of the Green's function at zero collision energy \cite{nishida2010confinement}, and $\tilde a(E)^{-1}=a_{3D}^{-1}-\mu_{KRb}\,r_{3D}E/\hbar^2$ is the energy-dependent scattering length \cite{petrov2004,levinsen2009}, with $a_{3D}$ and $r_{3D}$ the scattering length and the effective range in free space, respectively, and $\hat E_c$ the collision energy operator provided by
\begin{equation}
\hat E_c = \hbar\omega_{yz}^{(K)} - \left[-\frac{\hbar^2\grad_{\!\bm\rho}^2}{2(m_K+m_{Rb})}
+ \frac{m_K}{2}\omega_{yz}^{(K)2}\bm\rho^2\right].
\end{equation}
Once $\chi_m(\rho)$ is numerically obtained, the effective scattering length defined with the scattering amplitude being $\lim_{k\to0}f_m(k)\propto-a_{eff}^{(m)}k^{2m}$ follows from
\begin{align}
a_{eff}^{(m)} = \sqrt\frac{m_{Rb}}{\mu_{KRb}}
\int\!d^{2\!}\bm\rho'\rho'^me^{-\rho'^2/2l_{yz}^2}\chi_m(\rho').
\end{align}
The resulting $a_{eff}^{(m)}/l_{yz}^{2m+1}$ is shown in Fig.\ \ref{Fig1th} for $m=0$ and $m=1$ by utilising the free-space scattering length as a function of the external magnetic field from \cite{simoni2008near} and the free-space effective range fixed at 168.37 $a_0$ from \cite{lamporesi2010scattering}. We note that it is necessary to include the effective range correction for quantitative predictions. Confinement-induced resonances appear when the effective scattering length diverges, whose positions are also indicated by the vertical lines in Fig.\ \ref{Fig2} a).

To detect the interdimensional Feshbach resonances we look at the survival population of the K atoms as a function of the external magnetic field. As described earlier, we first set the magnetic field to the target value, and we then immerse the K atom in the Rb cloud keeping the magnetic field constant. After a fixed interaction time, we detect how many K atoms are left. Given that on average we trap $\simeq$ 1 atom of K, we expect three-body losses only through the KRbRb channel. This is verified by the exponential decay of $N_K$ as a function of the interaction time (see also in the next section). Our results for a Rb cloud at 1 $\mu$K, $\lambda_{SST}=790.025$ nm, and interaction time 50 ms are shown in Fig.\ \ref{Fig2} a). This set of parameters implies  1D-3D mixed dimensional scattering, since $k_BT/\hbar$ is lower than the radial trapping frequencies but significantly higher than the axial one. {Our data show several loss features between 20 and 45 G, as expected in this scattering regime. To quantitatively analyze the data, we perform Gaussian fits of those features whose amplitude is above the experimental errorbars, i.e., when the signal to noise ratio is above 1, and that are broader than twice our scan resolution, which is 0.25 G \cite{SuppMat}. We identify 8 features, whose fitted centres are reported in Fig. \ref{Fig2} b) together with the values predicted by the theory, grouped according to the $m$ and $n$ indices. The widths of the resonances obtained with the Gaussian fits are reported as `errorbars'. The spacing of the observed loss features clearly mirrors the one of the theoretical values and, in particular, the presence of two series corresponding to $m=0$ and $m=1$ is apparent. For what concerns the position of the resonances, the agreement between theory and experiment is very good. However, we were not able to detect loss features in correspondence to the $\{m=1, n=1\}$ and $\{m=0, n=4\}$ resonances. The strength of the resonances, that is responsible for the width and depth of the loss features, is expected to decrease as $m$ and $n$ increase. However, as also reported in other systems \cite{haller2010confinement,lamporesi2010scattering,capecchi2022observation}, our data do not show the expected behaviour. The positions of three-body loss features are affected by complicated dynamics between atoms and molecules. The center, width and the shape of the experimental curves can change with temperature, can be affected by other close resonances, or by the presence of Efimov resonances. A description of these processes is outside the scope of this work. \resp{Another effect that could limit the resolution of our Feshbach spectrosopy is the fact that during the immersion, the K atom experiences a different elastic collision rate for different values of the magnetic field. This implies that the thermalisation time with the bath is not uniform across the magnetic field scan, and in some cases the thermalisation time could even be longer than the lifetime of the K atom. This effect could be mitigated by first immersing the K atom at a value of the magnetic field that enables thermalisation, and then sweeping the magnetic field.}

We found that the agreement with the theoretical predictions is good if we utilize the free-space scattering length calculated in \cite{simoni2008near}. 
There is worse agreement if we instead utilize the free-space scattering length inferred in \cite{weber2008association}. Since there are many features in our system that are significantly narrower than in the free-space case, our measurements suggest that the position of the free-space resonance is closer to the one indicated in \cite{simoni2008near}. Remarkably, we can clearly identify both s-wave [orange lines in Fig.~3 a), points in Fig.~3 b)] and p-wave (cyan lines and points) resonances induced by the s-wave interaction. 

By lowering the temperature of the bath until $k_BT/\hbar\leq$ min$\{\omega^{(K)}\}=\omega^{(K)}_x$, we can bring the K atom close to its quantum ground state, and collisionally freeze all degrees of freedom. This is of particular interest for the implementation of several quantum information protocols. In this configuration, the scattering between the quantum impurity and the bath is in the 0D-3D regime. The anisotropy of the tweezer confinement is reflected in the scattering, so that resonances are approximately expected to appear when 
\begin{eqnarray}
   &&\left(1+|m|+2n\right)\sqrt{\frac{m_K}{m_K+m_{Rb}}}\hbar\omega^{(K)}_{yz} \nonumber \\
   &&+\left(\frac{1}{2}+n_x\right)\sqrt{\frac{m_K}{m_K+m_{Rb}}}\hbar\omega^{(K)}_{x}-|E_b| \nonumber \\ 
   &&=\hbar\omega^{(K)}_{yz}+\frac{\hbar\omega^{(K)}_{x}}{2}
   \label{eqbinding2}
\end{eqnarray}
with an integer $n_x\geq0$ labelling the new series of resonances caused by the confinement along the weak $x$ axis. These $n_x$ series are in addition to the resonances discussed in the 1D-3D case, that are still present. It is apparent from Eq.\ (\ref{eqbinding2}) that the spacing and therefore the width and strength of the new series of resonances is approximately a factor $\omega^{(K)}_x/\omega^{(K)}_{xy}$ smaller than for the 1D-3D resonances. Our experimental results for a Rb bath at 100 nK with trapping frequencies $\omega^{(Rb)}=2\pi\times(42,157,142)$ Hz, $\lambda_{SST}=$ 790.035 nm, and interaction time 150 ms are shown in Fig.\ \ref{Fig22} a). \resp{This value of $\lambda_{SST}$ generates a more repulsive potential for Rb than the one used for the 1D-3D scan. This is to compensate for the higher density of the Rb cloud at lower temperature, than generates higher collision rates.} We observe that, as expected, the position of the loss features is approximately unchanged. However, their shape is substantially different due to the additional $n_x$ resonances. These are so narrow and finely spaced that their net effect is to broaden the larger resonances and skew them towards lower magnetic field values. We show this effect in Fig.\ \ref{Fig22} b), which is an enlargement of the grey shaded area of Fig.\ \ref{Fig22} a). \resp{Such finely spaced resonances make it extremely challenging to finely control the interactions in the 0D-3D regime using an external magnetic field.}   

 \begin{figure*}
 \centering
 \includegraphics[width=0.99\textwidth]{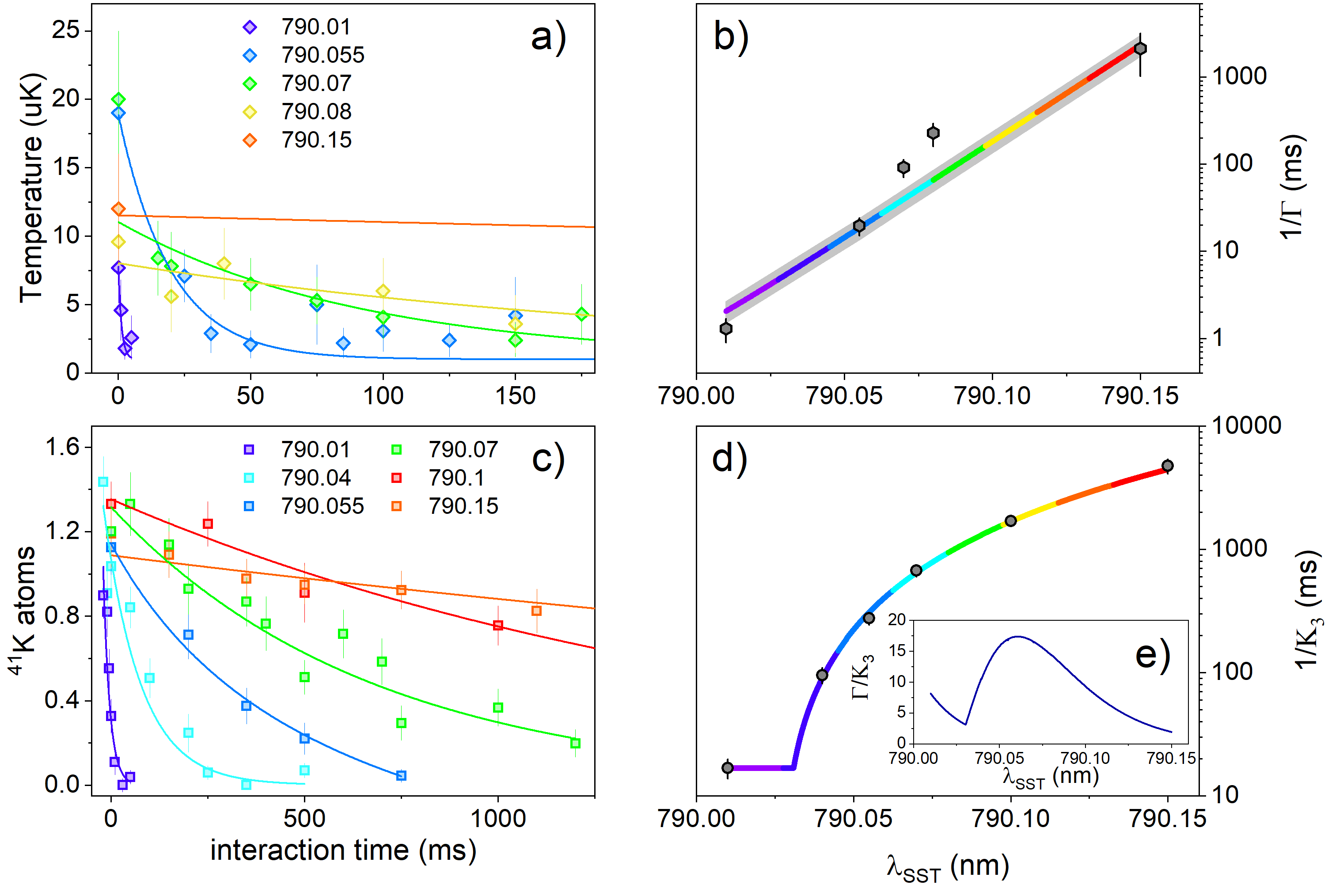}
 \caption{a) Temperature of the K atom in the SST as a function of the interaction time with the Rb bath, for different wavelengths of the SST. The magnetic field is set to 30 G, the starting temperature of the K atoms is $\simeq$ 20 $\mu$K, the bath is composed of 10$^{5}$ Rb atoms at 1 $\mu$K. The K temperatures and relative errorbars have been estimated with release-recapture thermometry and utilizing a bayesian estimator \cite{glatthard2022optimal}. Each data point corresponds to 180 experimental repetitions. The solid lines are exponential fits to the data with fixed asymptotic temperature of 1 $\mu$K. Most of the data points for $\lambda_{SST}=790.15$ nm are outside the range displayed. {For the full set of data see \cite{SuppMat}.} b) Thermalization time $1/\Gamma$ for K atoms  as a function of the SST wavelength. The values of $1/\Gamma$ and the errorbars are resulting from fitting the curves in panel a) with exponential decays. The solid line is a fit using Eq.\ (\ref{eqx}) with our experimental parameters, leaving $\xi$ as free parameter. The grey shaded area accounts for the error in the determination of $\xi$. c) Number of K atoms trapped in the SST as a function of the interaction time with a bath of 10$^{5}$ Rb atoms at 1 $\mu$K at 30 G. Each data point corresponds to 30 experimental repetitions. The errorbars are the standard deviation of the mean. The solid lines are exponential fits to the data with 0 as the asymptote. d) Three-body loss rate $\/K_3$ for the K atoms in the SST  as a function of $\lambda_{SST}$. The values of $1/K_3$ and the errorbars result from fitting the curves in panel c) with exponential decays. The solid line is a fit to the data with the maximum between a parabolic function and a lower threshold.  e) Ratio between $\Gamma$ and $K_3$ evaluated as the ratio between the solid curves in b) and d).}
 \label{Fig4}
 \end{figure*}

\section{Controlling the interactions with the species-selective tweezer}

The use of Feshbach resonances enables us to tune the scattering length between the quantum impurity and the atoms in the bath, and even generating resonant p-wave scattering from the underlying s-wave collisions. It is however challenging to precisely tune the interactions, especially in the 0D-3D regime, because of the many close and narrow resonances. This in turn also implies that the `zero-crossings' of the scattering length are very narrow, making it even more difficult to tune the interactions to zero. The ability to completely isolate the quantum impurity from the bath is particularly desirable, as it allows one to operate on the two subsystems independently. This is necessary for example to implement quantum engines, where quantum adiabatic transformations require decoupling the working fluid and the bath \cite{quan2007quantum, barontini2019ultra}. As anticipated earlier, the use of the SST provides us with an additional tool to progressively screen the K atom from the atoms in the bath. As shown in Fig.\ \ref{Fig5} a) and Fig.\ \ref{Fig5} b), by changing the wavelength of the tweezer light across $\lambda_0$, one can tune the tweezer potential for the Rb atoms from attractive to transparent to repulsive. Specifically, the SST potential becomes attractive (repulsive) for Rb atoms below (above) the tune-out wavelength. The potential for the K atom is instead essentially unaffected in our range of operations, see Fig.\ \ref{Fig5} a). This means that with the SST wavelength we can control the density of Rb atoms in the tweezer volume, and in turn control the scattering rate between Rb and K.

In Fig.\ \ref{Fig5} c) we show the number of remaining K atoms for a series of lower resolution scans of the magnetic field in the [20-35] G region, for different values of the SST wavelength. The more we tune the SST above $\lambda_0$, the higher the average number of K atoms remaining in the tweezer, signalling a lower overall interaction strength. We observe that when changing the wavelength, the position of the features is essentially unchanged, confirming that the change in wavelength does not affect the K potential. We instead observe that the width of the features in general increases for lower wavelength. This is a saturation effect due to the overall lower number of remaining K atoms. By increasing the wavelength the contrast of the loss features first increases, and then decreases again when the scattering rate vanishes. For $\lambda_{SST}\geq790.1$ nm, we are not able to detect any loss features at this interaction time. 

To characterize the way the quantum impurity interacts with the thermal bath, we measure the thermalization rate for different values of the SST wavelength for a fixed value of the magnetic field of 30 G. For this set of measurements $T_{Rb}\simeq1$ $\mu$K. As described earlier, before being immersed in the Rb cloud, the temperature of the K atoms in the tweezer is $T_K\simeq20$ $\mu$K. As reported in Fig.\ \ref{Fig4} a), we observe that, once immersed, the temperature of the K atoms decreases exponentially until it reaches the temperature of the bath. %\footnote{{Note that at $t=0$ in Fig. 6 a), the temperature of the K atoms is lower than 20 $\mu$K. This is because the K atom starts to interact with the Rb cloud in the 50 ms needed to immerse it. The temperature at $t=0$ slightly varies for different datasets because this also depends on the positioning of the SST with respect to the Rb cloud, which has about 5 $\mu$m uncertainty}}. 
This is because 
\begin{equation}
\frac{dT_K}{dt}=-(T_K-T_{Rb})\Gamma,
\label{eqz}
\end{equation}
with $\Gamma=\Gamma_{KRb}/\xi$ the thermalization rate, where  
\begin{equation}
    \Gamma_{KRb}=\sigma_{KRb}\sqrt{\frac{8k_B}{\pi}\left(\frac{T_K}{m_K}+\frac{T_{Rb}}{m_{Rb}}\right)   }\int n_K n_{Rb} d^3x
    \label{eqx}
\end{equation}
is the elastic scattering rate \cite{guttridge2017interspecies}, $\sigma_{KRb}=4\pi a_{eff}^{(0)2}$ the scattering cross section, and $\xi$ the number of collisions needed to thermalize, which is a function of the initial $T_K$ and the bath temperature $T_{Rb}$ \cite{lewenstein1995master,papenbrock2002rate,sawant2021thermalization}. To estimate $\xi$, we utilize Monte Carlo simulations that take into account the quantization of the degrees of freedom of the K atom \cite{sawant2021thermalization}. For our set of parameters we obtain $\xi\simeq12$. In Fig.\ \ref{Fig4} b) we report the thermalization time constant $1/\Gamma$ obtained from exponential fits of the curves in  Fig.\ \ref{Fig4} a). We observe that $1/\Gamma$ increases exponentially with $\lambda_{SST}$. This is expected, because the integral in Eq.\ (\ref{eqx}) decreases exponentially with $\lambda_{SST}$ \cite{SuppMat}, while $\xi$ remains constant since the initial and final temperatures do not vary. We fit our data with Eq.\ (\ref{eqx}), calculated using the theoretical value of $a_{eff}^{(0)}$ at 30 G, which is $0.7l_{yz}$, leaving $\xi$ as free parameter. With this procedure we find that $\xi=17.5\pm5$, which agrees with the value calculated with the Monte Carlo simulations.

For the same set of parameters, together with thermalization, we observe losses of K atoms as they interact with the Rb bath. We have indeed used these losses to detect the resonances in Fig.\ \ref{Fig2} and Fig.\ \ref{Fig5}. In our system, three-body losses could occur through two channels, namely KKRb and KRbRb. For the K density, three-body losses can therefore be described by the following rate equation:
\begin{equation}
    \frac{dn_K}{dt}=-2\alpha_{KKRb}n_K^2\tilde{n}_{Rb}-\alpha_{KRbRb}n_K\tilde{n}_{Rb}^2,
    \label{eqy}
\end{equation}
where $\alpha_j$ is the collision rate per unit of volume in the $j$ channel, that in general has a powerlaw dependence on the scattering length, and $\tilde{n}_{Rb}$ is the local density of Rb in the tweezer volume \cite{SuppMat}. Since in our system we have only $\simeq1$ K atom, we expect the first term in Eq.\ (\ref{eqy}) to be negligible. This can be verified by fitting the decay of the number of K atoms shown in Fig.\ \ref{Fig4} c) with the non-exponential solution of $\dot{n}_K=-2\alpha_{KKRb}n_K^2n_{Rb}$. In our case such a procedure delivers unphysical results. Our data are instead well fitted by an exponential decay, for which we conclude that the second term in Eq.\ (\ref{eqy}) is the dominant contribution for our atom losses. This is further confirmed by the quadratic dependence of the decay constant $1/K_3=1/(\alpha_{KRbRb}\tilde{n}_{Rb}^2)$ on $\lambda_{SST}$, shown in Fig.\ \ref{Fig4} d). Indeed, across $\lambda_0$, the Rb potential grows approximately linearly, therefore $\tilde{n}_{Rb}$ decreases linearly, implying the quadratic growth of $1/K_3$. The quadratic dependence breaks down for lower values of $\lambda_{SST}$, closer to $\lambda_0$, where the three-body loss rate $K_3$ saturates. This contributes to ensuring that, in our range of operation, the thermalization rate is always larger than the three-body loss rate, up to more than one order of magnitude, as shown in Fig.\ \ref{Fig4} e). This is a crucial ingredient for achieving thermalization. For this set of parameters, the optimal wavelength for efficient thermalization is therefore $\simeq790.06$ nm, corresponding to thermalization time scales of tens of ms. If faster thermalization rates are desired, the optimal conditions are achieved by further blue detuning $\lambda_{SST}$, exploiting the saturation of the three-body loss rate $K_3$. {This regime is of particular interest for the generation of trapped atomic dressed states like polarons, where strong repulsive interactions could counteract the strong attraction from the SST.}

In summary, in the region of parameters that we have investigated, we demonstrated the ability to use the wavelength of the SST to vary the thermalization rate of the quantum impurity system in the $\Gamma\simeq(1\rightarrow5\times10^{-4})$ s$^{-1}$ range and, accordingly, the three-body collision rate in the $K_3\simeq(0.05\rightarrow2\times10^{-4})$ s$^{-1}$ range. This translates as the ability to tune the screening of the single atom quantum impurity over a broad range, up to completely decoupling the single atom and the bath. 

\section{Conclusions}

In this work, we have demonstrated the implementation of a cold atom quantum impurity system by trapping a single K atom in an optical tweezer and immersing it in a cloud of ultracold Rb atoms. We have developed an experimental procedure that allows us to engineer all the parameters of the system combining optical tweezer techniques with standard ultracold gas methods. We achieved selective manipulation of the quantum impurity and the bath by realizing the optical tweezer with laser light tuned close to $\lambda_0$. We have shown that with this experimental architecture it is possible to fully control the interactions between the impurity and the bath acting on three parameters: the external magnetic field, the dimensionality of the motion of the K atom, and the wavelength of the SST. 

We performed Feshbach spectroscopy measuring the survival probability of the K atom in the bath as a function of the external magnetic field, detecting several loss features in the 1D-3D regime. We have used a theory including all the experimental parameters, that allowed us to associate the observed loss features to inter-dimensional Feshbach resonances. We detected two series of resonances, namely s-wave and p-wave, stemming from the underlying free-space s-wave Feshbach resonance. By lowering the temperature, we changed the dimensionality of the motion of the K atom from 1D to 0D. That in turn resulted in the appearance of additional narrow resonances linked to the finer energy level spacing along the weak axis of the SST. As a result, our measurements and characterization enable the control of the $H_{IB}$ term in Eq.\ (1) using the external magnetic field and the dimensionality of the system.    

We have shown that the wavelength of the SST can be used to screen the K atom from the atoms in the bath, down to almost completely decoupling them. Above $\lambda_0$, the SST becomes increasingly repulsive for the Rb atoms in the bath, and we demonstrated that this leads to an exponential decrease of the elastic scattering rate $\Gamma$. We have shown that the three-body loss rate $K_3$ instead scales quadratically, and that it is always lower than $\Gamma$, allowing the achievement of thermalization in our system over the whole range of parameters explored. We finally discussed that due to the different scaling of $K_3$ and $\Gamma$, there exists an optimal value of the SST wavelength where the $\Gamma/K_3$ ratio is maximum and non-hermitian effects are minimized. The tuning of the wavelength of the SST enables a greater degree of control over $H_{IB}$, notably decoupling the quantum impurity and the bath.    

Our work provides crucial elements for the implementation of pristine quantum simulators of quantum impurity models, like the spin-boson model and the Anderson impurity model, where the coupling between the impurity and the bath can be controlled and engineered. Our experimental architecture \resp{complements the features of systems like \cite{xPhysRevA.82.042722,yyPhysRevLett.109.235301},} and lends itself to the engineering of open quantum systems, in particular to the realization of environmental dark states, and the manipulation and control of qubits using the environment \cite{verstraete2009quantum,muller2012engineered}. All this can have practical applications in quantum information protocols \cite{goold2016role,alicki2004thermodynamics,loss1998quantum}. Particularly appealing is the possibility of studying thermodynamics in the quantum regime, more specifically the implementation of single atom quantum engines \cite{barontini2019ultra,rossnagel2016single,bouton2021quantum}, where also the bath can be used as a quantum resource \cite{uzdin2015equivalence,scully2003extracting,campo2014more,quan2007quantum}, \resp{and  advanced techniques like shortcuts to equilibration/thermalization/adiabaticity can further suppress losses \cite{ffPhysRevLett.122.250402,fffPedram_2023,ffffboubakour2024dynamical}.} In this context, the quantification of the energy cost of quantum technologies is becoming a pressing issue \cite{auffeves2022quantum}. Our platform could help in the understanding of how to efficiently manage the energy consumption in the quantum regime by acting on the interactions between the quantum system and the environment. Finally, there is huge potential in extending our methods to tweezer arrays \cite{endres2016atom,barredo2016atom}, in which the interactions between the single trapped atoms and the common bath could be individually and collectively tuned. This could lead to the creation of environmental resilient many-body entangled states \cite{amico2008entanglement,abanin2019colloquium,aolita2015open}, the study of bath mediated interactions \cite{zell2009distance}, thermal machines reaching quantum supremacy \cite{jaramillo2016quantum}, and novel quantum computation architectures \cite{kaufman2021quantum}.

\textit{Acknowledgments.} We acknowledge fruitful discussion with L. Correa, J. Glatthard, J. Rubio, and the Atomic Quantum Systems group at the University of Birmingham. We thank A. Simoni for providing the free-space molecular binding energy and scattering length, R. Sawant for the assistance with the code to calculate the thermalization rates, and A. Deb and V. Guarrera for reading the manuscript. We acknowledge the use of computing power provided by the Advanced Research Computing services at the University of Birmingham. This work was partially supported by the Leverhulme Trust Research Project Grant UltraQuTe (grant number RGP-2018-266) and by JSPS KAKENHI Grant No.\ JP21K03384.

\bibliography{name}

\end{document}